\documentclass[aps,amsmath,amssymb,pra,showpacs,twocolumn]{revtex4}

\usepackage[german,english]{babel}
\usepackage{amssymb}
\usepackage{amsfonts}
\usepackage{graphics}

\begin{document}

\title{Tight and loose shapes in flat entangled dense polymers}

\author{Andreas Hanke}
\affiliation{Institut f{\"u}r Theoretische Physik, Universit{\"a}t
Stuttgart, Pfaffenwaldring 57, D-70550 Stuttgart, Germany}
\email{hanke@theo2.physik.uni-stuttgart.de}
\author{Ralf Metzler}
\affiliation{NORDITA, Blegdamsvej 17, DK-2100 Copenhagen {\O}, Denmark}
\email{metz@nordita.dk}
\author{Paul G. Dommersnes}
\affiliation{Institut Curie, 11 rue Pierre et Marie Curie, F-75231 Paris
Cedex 5, France}
\author{Yacov Kantor}
\affiliation{School of Physics and Astronomy, Raymond and Beverly
  Sackler Faculty of Exact Sciences, Tel Aviv University, Tel Aviv 69978, Israel}
\author{Mehran Kardar}
\affiliation{Department of Physics, Massachusetts Institute of Technology,
Cambridge, Massachusetts 02139, USA}

\date{\today}

\begin{abstract}
We investigate the effects of topological constraints (entanglements)
on two dimensional polymer loops in the dense phase, and at the
collapse transition ($\Theta$ point).
Previous studies have shown that in the dilute phase the entangled region 
becomes tight, and is thus localised on a small portion of the polymer.
We find that the entropic force favouring tightness is considerably
weaker in dense polymers.
While the simple figure-eight structure, created by a single crossing in the 
polymer loop, localises weakly, the trefoil knot and all other prime knots 
are loosely spread out over the entire chain.
In both the dense and $\Theta$ conditions, the uncontracted knot configuration
is the most likely shape within a scaling analysis. 
By contrast, a strongly localised figure-eight is the most likely
shape for dilute prime knots.
Our findings are compared to recent simulations.
\end{abstract}


\maketitle

\section{Introduction}

Knots, and topological constraints in general, play an important role
in macromolecular systems. 
In gels and rubbers, permanent entanglements strongly influence the 
equilibrium and relaxation properties \cite{degennes,DE86,ferry,treloar}.
Even single molecules with identical chemical structure but different
topology may exhibit different physical properties \cite{sauvage}.
Knots are also present in biological molecules: For example,
some proteins exhibit knotted configurations \cite{creighton}, 
and the active modification of DNA knots through energy-consuming 
enzymes (topoisomerases) poses interesting challenges to the 
issue of the knot detection \cite{alberts,knotdetect}. 
Experimentally, the observation of individual molecules by 
single molecule force spectroscopy has come of age \cite{force,opt};
these methods can be used to probe the mechanical behaviour 
of knotted biopolymers directly. 

Given a knot in a closed ring, an obvious question is 
whether, on average, the knot segregates into a small 
region in which all topological details are confined, and a 
large, simply connected segment; or whether it is loosely spread
over the entire chain.
In the dilute phase, it has been found by numerical evidence
that flattened (hence two-dimensional \cite{2D}) knots are 
localised, i.e. {\em tight} \cite{knots2d}.
In a previous 
study \cite{slili2d}, we developed an analytical approach, 
based on scaling results for polymer networks 
\cite{duplantier,networks}, which explains and quantifies the tightness
of any prime knot in flat (2D) dilute polymer loops. 
Some aspects of our results 
have been verified experimentally by means of a vibrated 
granular chain \cite{HDBE02}. 
It should be mentioned that DNA chains have experimentally
been flattened by adsorption on an adhesive membrane \cite{raedler}.
In 3D, the global topological constraints of a knot are hard to 
implement by analytical methods \cite{GK94}.
(A simpler topological invariant, the linking number of closed  
DNA rings, can be incorporated in the mapping to a field 
theory \cite{degennes} by means of suitable gauge fields \cite{MK97}.)
Consequently, the tightness or localisation of 3D knots has not been
conclusively characterised.
A number of phenomenological models 
and numerical studies support the localisation of simple knots \cite{knots3d}.
Tightness has also been found in 3D slip-linked polymer 
chains in the dilute phase \cite{paraknot}.
Conversely, delocalisation
has been predicted for more complicated knots \cite{feigel}. 

In many situations, however, polymer chains are not dilute.
Polymer melts, gels, or rubbers exhibit fairly high densities of 
chains, and the behaviour of an individual chain in such systems
is significantly different compared to the dilute phase 
\cite{degennes,DE86}. Similar considerations apply to biomolecules:
In bacteria, the gyration radius of the almost freely floating 
ring DNA is often larger than the cell radius itself. 
Moreover, under certain conditions, there is a non-negligible 
osmotic pressure due to vicinal layers of protein molecules, 
which tends to confine the DNA \cite{walter}.
In protein folding studies, globular proteins in their native 
state are often modelled as compact polymers on a lattice 
(see \cite{Garel03} for a recent review).

Given this motivation, in this work we consider self-avoiding 
{\em dense} polymers with permanent entanglements, complementing
our previous studies for the dilute phase \cite{slili2d,paraknot}.
A polymer is considered dense
if, on a lattice, the fraction $f$ of occupied sites has a finite
value $f > 0$. This can be realised by a single polymer
of total length $L$ inside a box of volume $V$ and taking the limit 
$L \to \infty$, $V \to \infty$ in such a way that $f = L/V$ stays 
finite \cite{Dup86_a,DS_a}.
Alternatively, dense polymers can be obtained in an infinite volume
through the action of an attractive force between monomers. 
Then, for temperatures $T$ below the transition
(Theta) temperature $\Theta$, the polymers collapse to a dense phase,
with a density  $f > 0$, which is a function of $T$ \cite{DS_a,OPB93}. 
For a dense polymer in $d$ dimensions, the exponent
$\nu$, defined by the radius of gyration 
$R_g \sim L^{\nu}$, is simply $\nu = 1 / d$. 
In 2D, the dense polymer phase for $0 < f < 1$
is related to a conformal field theory, and the resulting
scaling behaviour is known exactly \cite{Dup86_a,DS_a,JRS03}.
The limit $f = 1$ is realised in Hamiltonian paths, 
where a random walk visits every site of a given 
lattice exactly once. For some cases, their 
scaling behaviour is known exactly as well \cite{DD,kondev}.
Another way to make polymers collapse in 2D is to exert
a pressure on self-avoiding loops (2D vesicles), conjugate 
to their area, which results in double-walled, branched 
structures \cite{F89,OS}. Recently, the corresponding crossover 
scaling function from linear to branched polymers (lattice 
animals) has been obtained exactly \cite{Car01}. 
It is believed that the branched and dense polymer phases are 
different.

Here, we extend our previous scaling analysis of knots in
dilute polymers \cite{slili2d} to the dense phase, and at the $\Theta$ point.
The general conclusion is that the entropic mechanism for tightening 
of entanglements is considerably weaker in the denser regimes.
While the simple figure-eight structure is still tight in these phases,
the trefoil becomes loose, with the trends more pronounced 
in the dense phase.
Note that at the $\Theta$ point in 2D, the swelling exponent is
$\nu = 4/7$, implying a fractal dimension $1/\nu$  smaller than $d = 2$, 
and an asymptotic density of $f = 0$.
In a recent numerical study, Orlandini, Stella, and Vanderzande
\cite{OSV2002} (hereafter referred to as OSV)
investigate the tightness of the 2D projection 
of the trefoil knot, and find {\em delocalisation\/}, in
contrast to the strong localisation obtained in the dilute
case \cite{slili2d}.

The main focus of this work is thus the localisation of flat
entangled polymers in the dense phase and at the $\Theta$
point by means of scaling arguments, in analogy to the dilute 
case \cite{slili2d}.
When possible, we compare these results with our own numerical simulations,
as well as those by OSV. In section \ref{sec_loc} we 
first review the differentiation between tight (localised) 
and loose (delocalised)
segments in entangled polymers. In section \ref{sec_F8} we then consider the 
2D figure-eight structure (F8) and compare our scaling results 
with Monte Carlo simulations. In section \ref{sec_trefoil} we derive 
the scaling results for the 2D projection of the trefoil, and
compare them with the simulations by OSV. Section \ref{sec_con} 
contains our conclusions. In appendix \ref{sec_app} we compile 
some scaling results for general polymer networks. 
Finally, at the end of appendix \ref{sec_app}, we consider 
briefly a newly discovered phase of 2D dense polymers where 
the strict non-crossing condition is relaxed \cite{JRS03}.


\section{Tight and loose segments} 
\label{sec_loc}

\begin{widetext}

\begin{figure}
\unitlength=1cm
\begin{picture}(10,10.5)
\put(-6.2,-8.8){\includegraphics{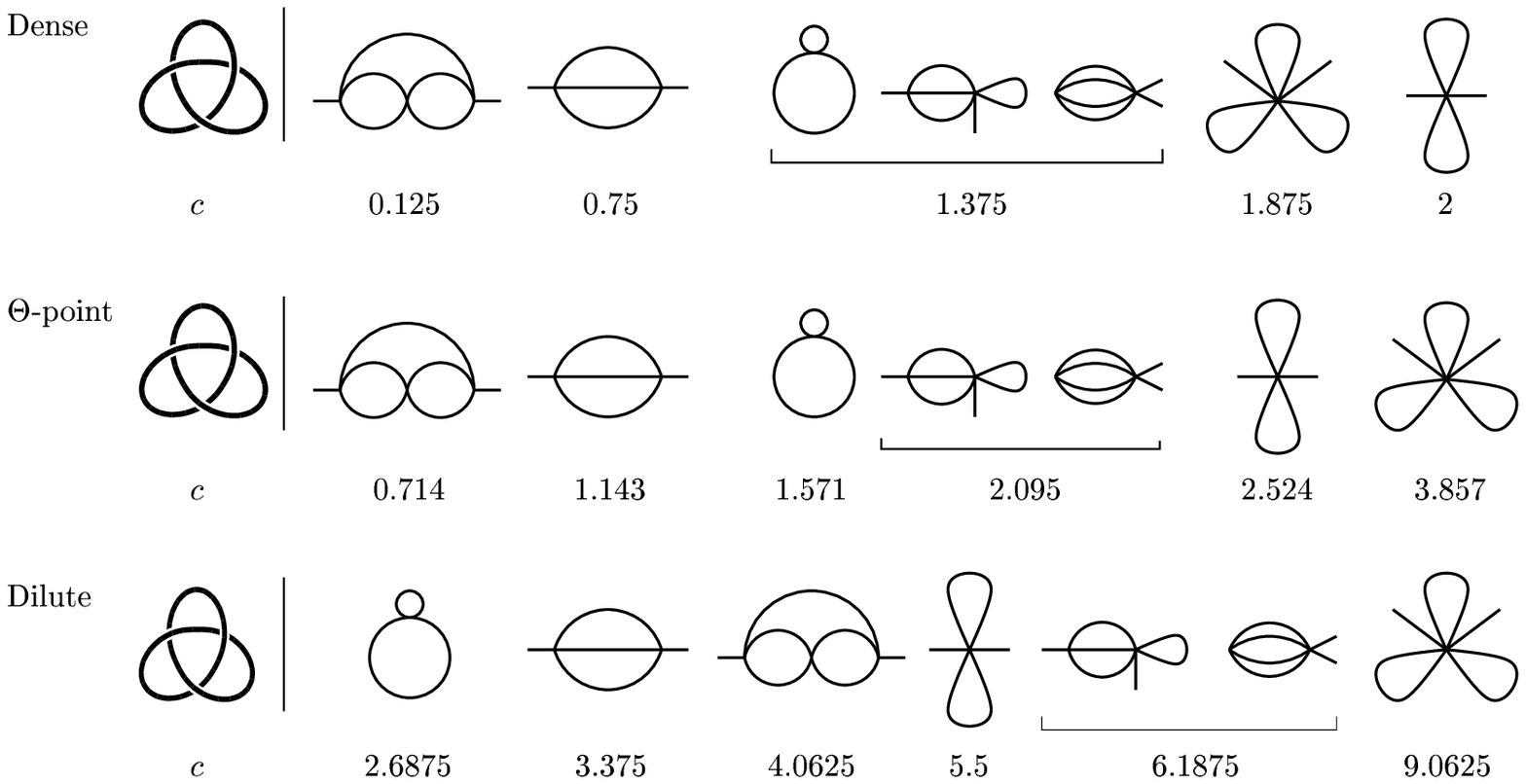}}
\end{picture}
\caption{``Contractions'' of the flat trefoil knot, arranged according
to higher scaling orders. The protruding legs stand for the remaining 
simply connected ring. Below the contractions, we list the
associated scaling exponent $c$ (see text). 
Both for dense polymers and polymers at the
$\Theta$ point, the leading shape is the original (uncontracted)
trefoil configuration on the very left.
For comparison, we also show the corresponding ordering for the
dilute phase \cite{slili2d}.
\label{fig31}}
\end{figure}

\end{widetext}

Consider an entangled polymer chain in 2D of length $L$, 
such as a simple flat, once-twisted ring with one crossing 
(called ``figure-eight'' (F8), see position III in figure
\ref{fig31} top row) or the 2D projection of 
the trefoil pressed flat against a surface by an external 
force (figure \ref{fig31}) \cite{2D}.
The orientations of the crossings are irrelevant, and can thus 
be considered as vertices with 4 outgoing legs. Thus,
each structure is mapped on a 2D polymer network with a number 
of vertices which are joined by ${\cal N}$ segments of variable 
lengths $\left\{s_i\right\}$ under the constraint 
$\sum_{i=1}^{\cal N} s_i = L$.
In the following we shall use the convention 
$s_1\leq s_2\leq \ldots\leq s_{\cal N}$.

Since we are interested in the tightness of such a network,
we define the size of the entangled region as 
$\ell = \sum_{i=1}^{{\cal N}-1} s_i$ so that the remaining 
(largest) segment  is $s_{\cal N} = L - \ell$.
Clearly, for the F8, ${\cal N} = 2$ and $\ell = s_1$, while for 
the trefoil ${\cal N} = 6$ and $\ell = \sum_{i=1}^5 s_i$.   
Note that the above definition  does not necessarily
imply that $\ell$ is small; however, if $\ell \ll L$, the 
structure assumes the form of a possibly multiply connected knot 
region of size $\ell$ and a large simple loop of size $L - \ell$. 
For a knot represented 
by the network ${\cal G}$, an important quantity is
its number of configurations $\omega_{\cal G}(\ell,L)$
for fixed $\ell$. 
In general, in the tight limit $\ell \ll L$, the
configuration number scales as a power law
\begin{equation} \label{power}
\omega_{\cal G}(\ell,L) \, \sim \, \ell^{- c_{\cal G}} \, \,
\quad (\ell \ll L) \, \, ,
\end{equation}
where the exponent $c_{\cal G}$ depends on the topology of the network
${\cal G}$ \footnote{In the remainder of this section, we drop the 
index ${\cal G}$ for ease of notation.}.

For given $\omega(\ell,L)$, various quantities of interest 
can be calculated. For example, the mean size of the knot 
region is given by
\footnote{The upper integration limit can be chosen as any finite
fraction of $L$ in principle; the choice of $L/2$ imposes no restriction
on generality.}
\begin{equation} \label{small}
\langle \ell \rangle = \int_a^{L/2} d\ell \, \ell \, p(\ell,L)\,\, ,
\end{equation}
with the (normalised) probability density function (PDF)
\begin{equation} \label{pdf}
p(\ell,L) \, = \, \omega(\ell,L) \bigg/ 
\int_a^{L/2} d\ell \, \omega(\ell,L) \, \, .
\end{equation}
We have introduced a short-distance cutoff $a$, set by the lattice 
constant, to control the non-integrable singularity at $\ell = 0$
which occurs in equation (\ref{small}) for $c>2$,
and in the denominator of equation (\ref{pdf}) for $c>1$.

\noindent
Thus, depending on the value of $c$,
three cases can be distinguished:

(i) $c < 1$: Both integrals are well-defined for $a = 0$, 
and one immediately obtains 
$\langle \ell \rangle \sim L$ due to dimensional considerations.
The knot region grows linearly with $L$ and is thus 
{\em delocalised\/}, i.e. spanning  the whole polymer.

(ii) $1 < c < 2$: To leading order in $a$, one finds 
$\langle \ell \rangle \sim a^{c-1} L^{2-c}$. The size of 
the knot region scales with $L$, but with an exponent $2-c < 1$,
and is thus {\em weakly localised\/}, i.e., 
$\langle \ell \rangle/L\to 0$ for $L \to \infty$.

(iii) $c > 2$:  To leading order in
$a$, one finds $\langle \ell \rangle \sim a$, i.e., the 
size of the knot region is independent of $L$ for  $L \to \infty$,
and is thus {\em strongly localised\/}.

In a more complicated network, such as for the trefoil in  
figure \ref{fig31}, several segments can be simultaneously tight.
We shall refer to the emerging structures as possible {\em shapes\/} 
(or contractions) of the original network. Each shape
corresponds again to a network ${\cal G}$.
The parameter $c$ not only determines the tightness 
of the knot region for given shape ${\cal G}$,
but also controls the {\em overall likelihood} $P_{\cal G}$, 
of different possible shapes.
The latter is obtained (up to  normalisation)
by integrating $\omega_{\cal G}(\ell, L)$
over all possible values of $\ell$, as in the denominator of 
equation~(\ref{pdf}). Depending on the values of $c$,
different cases can be distinguished:

(i) $c < 1$: The integral is convergent for small $\ell$ with a finite 
limit for $a \to 0$, and scales as $(a/L)^{c-1}$.
If there is a variety of shapes with $c<1$, the one with the lowest
value $c_m$ is the most likely, $P_{\cal G}$ for the others
scaling with a factor of $(a/L)^{c-c_m}$. 
This is expected, since networks with $c < 1$ are delocalised 
over the whole chain of length $L$.

(ii) $c>1$: The integral is divergent for small $\ell$, and is thus
dominated by the lower cutoff. To make sure that the knot
region is large enough to consider different vertices (crossings)
as separated on a scale much larger than $ a$, one should use a lower cutoff 
${\cal A}$ of intermediate length $a \ll {\cal A} \ll L$
\footnote{In a MC simulation, a larger ${\cal A}$ gives worse 
statistics since configurations with 
$\ell > {\cal A}$ become less frequent. In our case, we considered 
two vertices as separated if the segment between them was larger 
than 5 monomers. 
For the original trefoil this gives ${\cal A} \approx 25$ 
monomer lengths (5 monomers $\times$ 5 segments).}.
The relative probabilities for different shapes with $c>1$ now scale 
with a factor of  $(a/{\cal A})^c$.
Different shapes with $c > 1$ scale differently with the lower cutoff
${\cal A}$, but not differently with $L$. This reflects the 
fact that for $c > 1$ the probability $P_{\cal G}$ is dominated by 
the small $\ell$ behaviour of $\omega_{\cal G}(\ell, L)$.

In both cases $P_{\cal G}$ scales with a factor of $a^{c}$.
This is the {\em scaling order} of such a shape ${\cal G}$, 
as the likelihood of possible shapes with distinct values of 
$c$ can be ordered according to powers of $a^c$
with more dominant networks for smaller values of $c$
\cite{slili2d}.


\section{The 2D figure-eight:
Scaling analysis \& Simulations}
\label{sec_F8}

The most elementary entangled object in 2D is the F8, which 
consists of two loops of variable lengths
$\ell$ and $L - \ell$ 
(position III in figure \ref{fig31} top row).
The crossing point can be considered as a vertex with 4 outgoing legs, 
and the F8 corresponds to a network with 
${\cal N} = 2$ segments of lengths $s_1 = \ell$ and 
$s_2 = L - \ell$ \cite{2D}.
The number of configurations $\omega_{8}(\ell,L)$ 
for the F8 can be deduced from results for general polymer networks, 
obtained by Duplantier and coworkers \cite{Dup86_a,DS_a,DD,DS87},
which are compiled in appendix \ref{sec_app}.
In 2D it has the scaling form
\begin{equation} \label{scaling} 
\omega_{8}(\ell,L) \sim \omega_0(L) \, (L-\ell)^{\gamma_{8}} 
{\cal X}\left(\frac{\ell}{L-\ell} \right) \, .
\end{equation} 
As discussed in the previous section, the localisation of F8
is controlled by the limit 
of $\omega_8(\ell,L)$ for $\ell \to 0$, i.e., the behaviour of the 
scaling function ${\cal X}(x)$ for $x = \ell / (L-\ell) \to 0$. 
The latter can be determined by the following argument 
(see text below equation (59) in reference \cite{DS_a}): 
Clearly, for 
$\ell \ll L$, the big loop of length $L-\ell$ will behave 
like a simple ring, so that $\omega_{8}(\ell,L)$ 
should reduce to $\omega_0(L)$. This implies 
${\cal X}(x)\sim x^{\gamma_{8}}$ as $x \to 0$ and thus
\begin{equation} \label{sl_limit}
\omega_{8}(\ell,L) \sim \omega_{0}(L) \, \ell^{-c} \, 
\quad (\ell \ll L) \, . 
\end{equation} 
The value of the exponent $c$ for the 2D dense F8
(see appendix \ref{sec_app}) is
\begin{equation} \label{c8}
c = - \gamma_{8} = 11/8 = 1.375 \, \, ,
\end{equation}
implying that the smaller loop is {\em weakly localised\/}. This means
that the probability for the size of each loop is peaked 
at $\ell = 0$ and, by symmetry, at $\ell = L$. 
An analogous reasoning for the 2D F8 at the $\Theta$ point gives
\begin{equation} \label{c8_theta}
c = 11/7 = 1.571 \, \, ,
\end{equation}
i.e., in this case the smaller loop is also weakly localised.

\begin{figure}[h]
\unitlength=1cm
\begin{picture}(8,4.6)
\put(0,0){\includegraphics{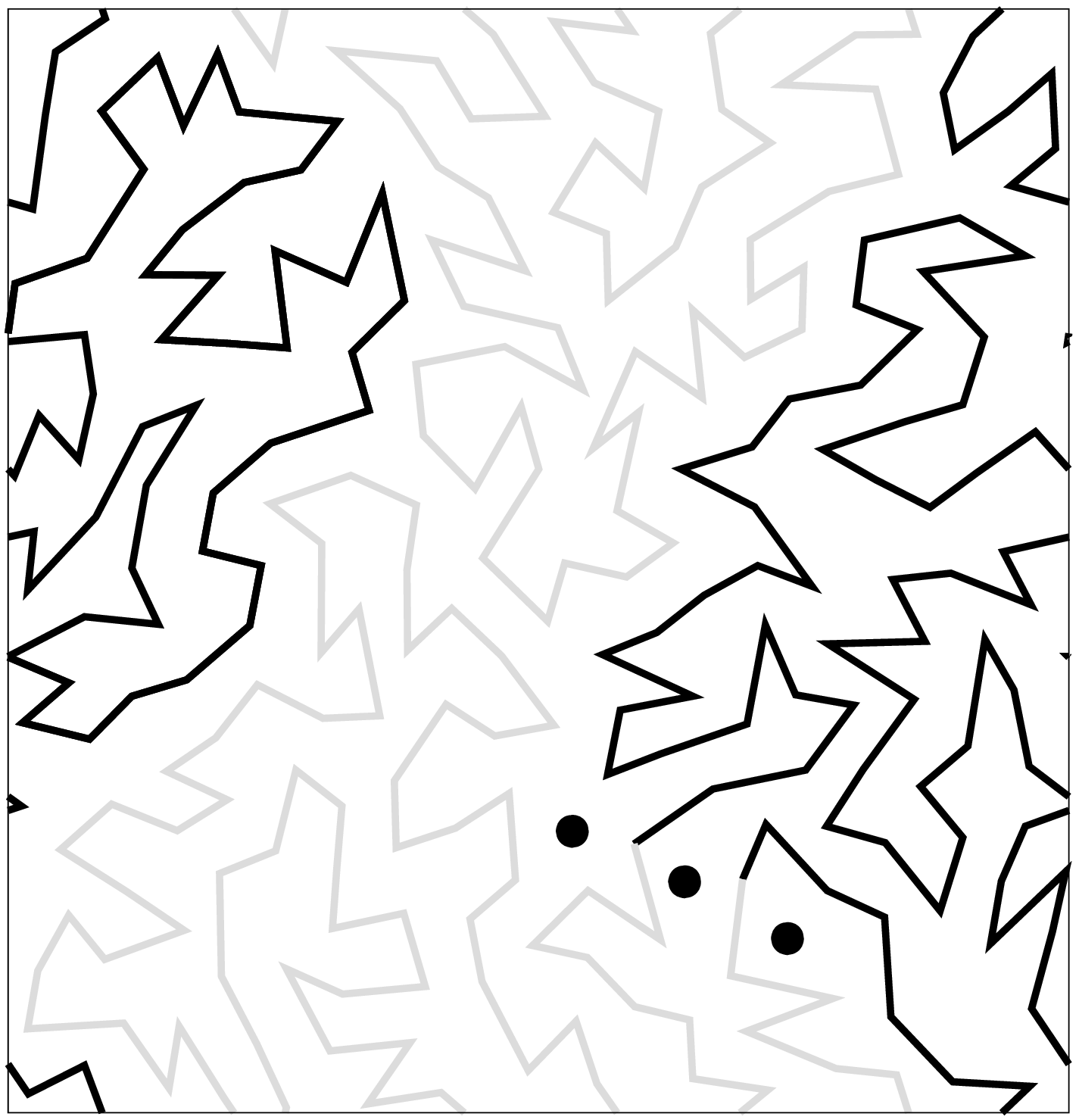}}
\put(4.4,0){\includegraphics{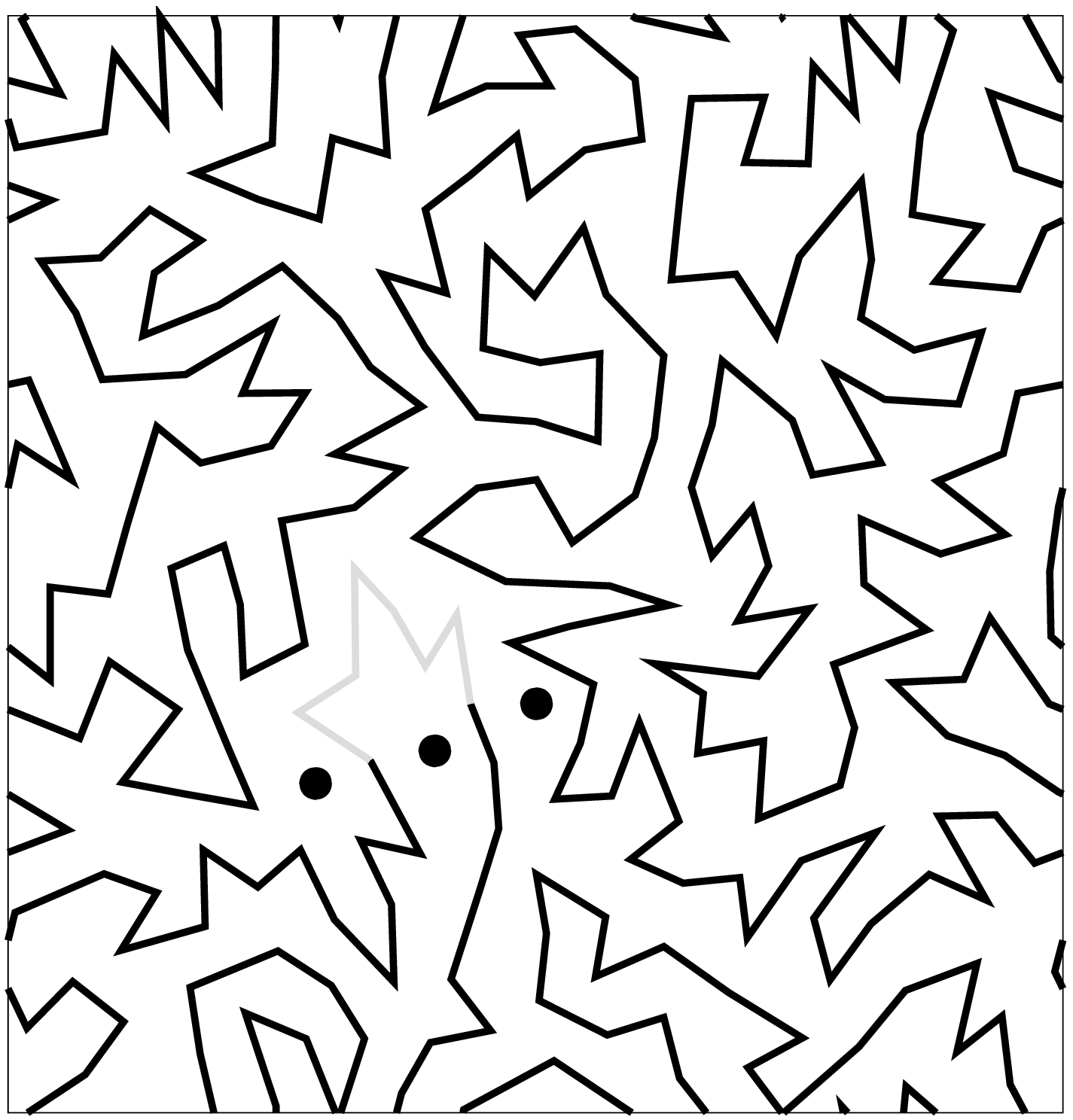}}
\end{picture}
\caption{Symmetric ($\ell=L/2=128$) initial configuration of a 2D dense
F8 (left), and its equilibrium configuration (right) with
periodic boundary conditions. The slip-link is
represented by the three (tethered) black dots.\label{fig-8}}
\end{figure}

\begin{figure}[h]
\unitlength=1cm
\begin{picture}(8,5.8)
\put(-0.5,-0.4){\includegraphics{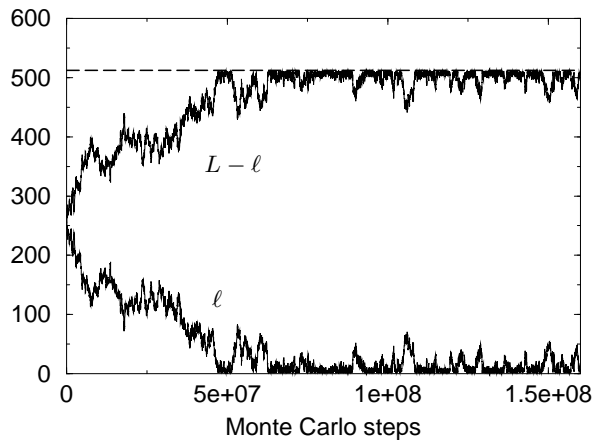}}
\put(3.1,3.6){$L-\ell$}
\put(3.2,1.8){$\ell$}
\end{picture}
\caption{The breaking of the initial symmetry between the two loops of the F8,
as  a function of MC steps at 55\% area coverage. 
\label{fig-8_1}}
\end{figure}

Figure \ref{fig-8} shows
the symmetric initial and a typical equilibrium configuration for periodic
boundary conditions obtained from Monte Carlo (MC) simulations, 
see below for details. 
In figure \ref{fig-8}, the lines represent the bonds (tethers) between the
monomers (beads, not shown here). The three black dots mark the locations
of the tethered beads forming the slip-link in 2D, by which we model
the crossing \cite{slili2d,paraknot}. 
The initial symmetric configuration soon gives way to a
configuration with $\ell\ll L$ on approaching equilibrium.
Figure \ref{fig-8_1} shows the development of this symmetry breaking as a 
function of the number of MC steps. 
We note, however,
that the fluctuations of the loop sizes in the ``stationary'' regime
appear to be larger in comparison to the dilute case studied in reference
\cite{paraknot}. We checked that for densities (area coverage) above
40\% the scaling behaviour becomes independent of the density.
The above simulation results correspond to a density of 55\%,
which is roughly half of the maximal possible density of 90\% 
(closed packed area coverage).

\begin{figure}
\unitlength=1cm
\begin{picture}(8,5.8)
\put(0,-0.9){\includegraphics{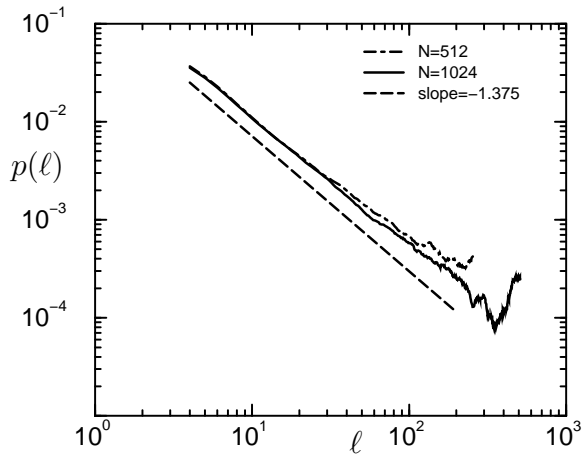}}
\put(0.5,3.7){{\large $p(\ell)$}}
\put(5,0){{\large $\ell$}}
\end{picture}
\caption{The loop size probability distribution  $p(\ell)$ at
$\rho=55\%$ area coverage, for the F8 with 512 (top line) and 
1024 monomers (solid line). 
The power law with the predicted exponent $c=1.375$ 
in equation (\ref{c8}) is indicated by the dashed line.
\label{fig3_1}}
\end{figure}

\noindent
The size distribution data is well fitted to a power law 
(for over 1.5 decades with 1024 monomers),
and the corresponding exponent  with 512 and 1024 
monomers in figure \ref{fig3_1} 
is in good agreement with the predicted value.

For our MC analysis, we used a hard core bead-and-tether chain, in
which self-crossings were prevented by keeping a maximum bead-to-bead
distance of 1.9 times the bead diameter, and a maximum step length of
0.27 times the bead diameter (cf. also \cite{2D}).
The slip-link was represented by three tethered beads enforcing the 
sliding pair contact such that the loops cannot fully retract, see 
figure \ref{fig-8}. To create the dense F8
initial condition, a free F8 is squeezed into a quadratic box with
hard walls.  This is achieved by starting off from the free F8,
surrounding it by a box, and turning on a force directed towards one
of the edges. Then, the opposite edge is moved towards the centre of
the box, and so on.  During these steps, the slip-link is locked,
i.e., the chain cannot slide through it, and the two loops are of
equal length during the entire preparation. Finally, when the
envisaged density is reached, the hard walls are replaced by periodic
boundary conditions, and the slip-link is unlocked. For all densities
studied we observe that one of the loops becomes much smaller than the
other one. We consider the chain as relaxed when the structure has
changed from symmetric ($s_1=s_2=N/2$) to asymmetric 
($s_1 \sim 1$, $s_2 \sim N$), then to symmetric and back again 
to asymmetric. After this we start to
sample the probability distribution for the loop lengths.  The
probability distribution with chain length $N=1024$ and density
$\rho=55\, \%$ required about $3 \times 10^9$ Monte Carlo steps 
($\sim 3 \times 10^{12}$ attempted moves).  
We note that an explicit simulation to obtain
the PDF $p(\ell)$ for more complicated structures than the
F8 was not possible within reasonable computation time with 
the MC algorithm we used.


\section{The 2D trefoil:
Scaling analysis \& Simulations} 
\label{sec_trefoil}

We now turn to the 2D projection of the trefoil (denoted ``3'', 
cf.~left part and position I in figure \ref{fig31} top row).
Each of the three crossings is replaced with a vertex of order
four, resulting in a network with ${\cal N} = 6$ segments 
of lengths $\left\{s_i\right\}$ 
and total length $L = \sum_{i=1}^{6} s_i$ \cite{2D}.
The size of the knot region is $\ell = \sum_{i=1}^5 s_i$
and $s_6 = L - \ell$. 
According to section \ref{sec_loc}, we need to know the behaviour 
of the configuration number $\omega_3(\ell,L)$ for $\ell \ll L$ 
in order to decide how $\langle \ell \rangle$ scales with $L$. 
For {\em fixed} segment lengths $\left\{s_i\right\}$ with $\ell \ll L$, the 
configuration number of the network can be derived by using 
equation (\ref{noc}) in appendix \ref{sec_app} and an analogous 
reasoning invoked to obtain equation (\ref{sl_limit}), resulting 
in
\begin{equation} \label{3_limit_prime} 
\omega'_{3} \sim \omega_{0}(L) \, \ell^{\gamma_3} \, 
{\cal W}\left( \frac{s_1}{\ell}, \ldots, \frac{s_4}{\ell} \right) \, , 
\quad \ell \ll L \, \, .
\end{equation} 
The prime on $\omega'_{3}$ indicates that the segment lengths 
$\left\{s_i\right\}$ are
fixed. However, since in our case the individual segments of the knot
region may exchange length with each other and only the total length 
$\ell$ of the knot region is fixed, we integrate $\omega'_{3}$ over 
all distributions of lengths $\left\{s_i\right\}$ under the constraint 
$\ell = \sum_{i=1}^5 s_i$ \cite{slili2d}. This yields the desired
number of configurations of the 2D trefoil with fixed $\ell \ll L$ as
\begin{equation} \label{3_limit}
\omega_{3}(\ell,L) \sim \omega_{0}(L) \, \ell^{-c}
\end{equation}
with $c = - \gamma_3 - m$, where $\gamma_3 = - 33/8$ from equation 
(\ref{nexp}) in appendix \ref{sec_app} (${\cal L} = 4$, $n_4 = 3$) 
and $m = 4$ is the number of independent integrations over chain 
segments. Thus, $c = 1/8 < 1$ which implies that the dense 2D 
trefoil is {\em delocalised}. 

The above analysis corresponds to the case for which all segments 
$\left\{s_i\right\}$ are
large compared with the short-distance cutoff $a$. Conversely, if 
some of the segments $\left\{s_i\right\}$ are of the order of $a$,
the vertices they join can no longer be resolved on macroscopic length scales,
but constitute a new, single vertex, possibly with more than four outgoing
legs. The corresponding {\em contractions\/} of the original 2D trefoil
thus represent different networks, each one with its
own topological exponent $\gamma$ and localisation exponent $c=-\gamma-m$
(where $m$ is the number of independent integrations, see above).
These contractions correspond to different {\em shapes\/},
which can be ordered according to their scaling order in $a$,
i.e., according to increasing powers of $c$ (cf.~section \ref{sec_loc}).

For {\em dilute\/} polymers, we have shown in this way that the leading 
scaling order is the F8 with $c = 43/16$, and that the original 
(uncontracted)
trefoil shape is only found at the third position \cite{slili2d}. 
For {\em dense\/} polymers, however, the present scaling results show
that both the original trefoil shape ($c = 1/8 < 1$, see above)
and position II ($c = 3/4 < 1$) are in fact {\em delocalised\/} 
(top row in figure \ref{fig31}). 
The F8 is only found at the third position and is weakly 
localised ($c = 11/8 > 1$, cf.~section \ref{sec_F8}).
Thus, in a MC simulation of the dense 2D trefoil, 
we predict that one mainly observes delocalised shapes corresponding 
to the original trefoil, and less frequently 
the other shapes of the hierarchy (top row) in figure \ref{fig31}.

These predictions are consistent with the numerical simulations of
Orlandini et al.~\cite{OSV2002} (OSV), who observe that the mean 
value of the second largest segment of the simulated 2D dense trefoil 
configurations grows linearly with $L$, and conjecture the same 
behaviour also for the other segments, corresponding to the 
delocalisation of the trefoil obtained above. However, we note
that the configuration shown in figure 1 of OSV corresponds in our
modelling to the shape at position II in figure \ref{fig31}, where we 
consider the two crossings to the right in figure 1 of OSV as one 
``molten'' vertex.

The same reasoning can be applied to the 2D trefoil in the 
$\Theta$ phase.
In this case we find that the leading shape is again the 
original (uncontracted) trefoil, with $c = 5/7 < 1$. This 
implies that the 2D trefoil is {\em delocalised\/} also at the 
$\Theta$ point. All other shapes are at least weakly localised,
and subdominant to the leading scaling order represented by the
original trefoil. The resulting hierarchy of shapes is shown 
in figure \ref{fig31} (middle row).

This finding is at variance with the simulation results of OSV
at the $\Theta$ point, who observe a behaviour of the simulated 
trefoil configurations similar to the F8, which is weakly
localised with $c = 11/7 > 1$, and is found only at the third 
position in the hierarchy of figure \ref{fig31}.


\section{Conclusions}
\label{sec_con}

We presented a scaling analysis for the 2D figure-eight structure 
(F8) and the 2D projection of the trefoil knot at equilibrium, 
both in the dense phase and at the $\Theta$ point \cite{2D}. 
Figure \ref{fig31} shows the hierarchy of contractions of 
the original trefoil, arranged according to scaling order. 
The F8 structure represents the leading order of the hierarchy in
the dilute phase \cite{slili2d}, but does not play a special role in 
the other states.
Thus we conclude that the 2D trefoil is {\em delocalised\/} both in the 
dense phase and at the $\Theta$ point, in contrast to its localisation 
in the dilute case \cite{slili2d}. 
The delocalisation of the flat trefoil in the dense phase has been
observed in the simulations by Orlandini et al.~\cite{OSV2002} (OSV).
However, their observation of behaviour corresponding to a weakly localised
F8 at the $\Theta$ transition contradicts our results.

If the chain's topology is that of a F8, 
then one of the loops is predicted
to be tight in all cases, although the localisation exponent 
depends on the polymer phase.
We explicitly verified the localisation exponent of $c = 11/8$ 
by a MC study of a F8 in the dense phase.
(We employed periodic boundary conditions for this simulations to avoid
any potential problems associated with surface effects.)

In 2D, the scaling analysis can be readily extended to 
general prime knots. Indeed, the minimal 2D projection of any 
prime knot can be mapped on a network ${\cal G}$, for which
one can calculate the corresponding exponent $c$ in a similar way 
as before. Using the Euler relations
$2 {\cal N} = \sum_{N \ge 1} N n_N$ and
${\cal L} = \sum_{N \ge 1} \frac{1}{2} (N-2) n_N + 1$,
we find
\begin{equation} \label{euler}
c_{\cal G}= 2 + \sum_{N\ge 4}n_N\left[\frac{N}{2}(d\nu-1)+
(|\sigma_N|-d\nu)\right] \, .
\end{equation}
Both for dense polymers and polymers at the $\Theta$ point in 2D 
one has $d \nu \ge 1$ and $|\sigma_N|$ increases with $N$, 
so that the term in the square brackets in the above expression
increases with $N$ as well. For a fixed number ${\cal V} = \sum _{N\ge 4}n_N$
of vertices this implies that $c_{\cal G}$ is minimal if only vertices
with four outgoing legs are present. Using this and the fact that for
$N = 4$ the term in the square brackets is negative
\footnote{Conversely, for dilute polymers this bracket is positive.
This is the reason why the order of the first 3 contractions in each
row of figure \ref{fig31} (which have only vertices with four outgoing 
legs) for dilute polymers is reverse compared to dense and 
$\Theta$ polymers.},
we conclude that
$c_{\cal G}$ is minimal if the number of such 4-vertices is maximal. This
implies that the leading scaling order of the 2D projection of {\em any} 
prime knot is the original (uncontracted) configuration, for which 
the above vertices represent the crossings, and this configuration will
be {\em delocalised\/} (since the 2D trefoil configuration is already 
delocalised).

We thank J. Cardy and U. Seifert for helpful discussions. This work 
was supported by the National Science Foundation (DMR-01-18213) and 
the US-Israel Binational Science Foundation (BSF) grant No.~1999-007. 
RM acknowledges partial support from the Deutsche 
For\-schungs\-ge\-mein\-schaft.


\begin{appendix}

\section{Scaling in general polymer networks}
\label{sec_app}

In this appendix we review the scaling results for polymer 
networks in the dense phase in 2D \cite{Dup86_a,DS_a,OPB93,DD} 
and at the $\Theta$ transition \cite{DS87}.

\begin{figure}
\unitlength=1cm
\begin{picture}(8,4.6)
\put(-2.4,-21.5){\includegraphics{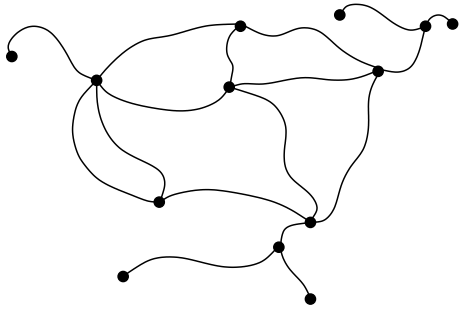}}
\end{picture}
\caption{Polymer network ${\cal G}$ with vertices ($\bullet$) of different
order ($n_1=5$, $n_3=4$, $n_4=3$, $n_5=1$).
\label{netw}}
\end{figure}

A general polymer network 
${\cal G}$, like the one depicted in Fig.~\ref{netw}, consists of a 
number of vertices which are joined by ${\cal N}$ chain segments
of total length $L$.
First, consider the dense phase in 2D. If all segments have equal 
length $s$ and $L = {\cal N} s$, the configuration number 
$\omega_{\cal G}$ of such a network scales as 
\cite{Dup86_a,DS_a}
\footnote{Note that due to the factor $\omega_0(L)$ the exponent of 
$s$ is $\gamma_{\cal G}$, and not $\gamma_{\cal G}-1$ like in the 
expressions used in the dilute and $\Theta$ phases,
for which  $\omega_0(L) \sim L^{- d \nu}$. However,
for 2D dense polymers one has $d \nu = 1$, so that both definitions of 
$\gamma_{\cal G}$ are equivalent, cf.~section 3 in reference \cite{DS_a}.}
\begin{equation} \label{noc_md}
\omega_{\cal G}(s) \sim \omega_0(L) \, s^{\gamma_{\cal G}} \, \, ,
\end{equation}
where $\omega_0(L)$ is the configuration number of a simple ring of 
length $L$. For dense polymers, and in contrast to the dilute and 
$\Theta$ phases, $\omega_0(L)$ (and thus $\omega_{\cal G}$) 
depends on the boundary conditions and even on the shape of the system 
\cite{DS_a,OPB93}. For example, for periodic boundary conditions (which 
we focus on in this study) corresponding to a 2D torus, one finds
$\omega_0(L) \sim \mu^L \, L^{\Psi - 1}$ with a connectivity
constant $\mu$ and $\Psi = 1$ \cite{DS_a}. However, the network 
exponent 
\begin{equation} \label{nexp}
\gamma_{\cal G} = 1 - {\cal L} + \sum_{N\ge 1}n_N\sigma_N
\end{equation}
is {\em universal\/} and depends only on the topology of the
network by the number ${\cal L}$ of independent loops, and by
the number $n_N$ of vertices of order $N$ with vertex exponents 
$\sigma_N = (4 - N^2)/32$ \cite{Dup86_a,DS_a}. For a linear chain,
the corresponding exponent $\gamma_{\rm lin} = 19/16$ has been 
verified by numerical simulations \cite{DS_a,GH95}. For a network 
made up of different segment lengths $\left\{s_i\right\}$ of total 
length $L = \sum_{i=1}^{\cal N} s_i$, equation (\ref{noc_md}) 
generalises to (cf.~section 4 in reference \cite{DS_a})
\begin{equation} \label{noc}
\omega_{\cal G}(s_1, \ldots, s_{\cal N}) 
\, \sim \, \omega_0(L) \, s_{\cal N}^{\gamma_{\cal G}} \,
{\cal Y}_{\cal G}\left(\frac{s_1}{
s_{\cal N}},\ldots,\frac{s_{{\cal N}-1}}{s_{\cal N}}
\right) \,  ,
\end{equation}
which involves the scaling function ${\cal Y}_{\cal G}$.

For polymers in an infinite volume and endowed with an attractive
interaction between neighbouring monomers, a different scaling behaviour 
emerges if the system is not below but right at the $\Theta$ point 
\cite{DS87}. In this case the number of configurations of a general 
network ${\cal G}$ is given by
\begin{equation} \label{noc_theta}
\overline{\omega}_{\cal G}(s_1, \ldots, s_{\cal N}) 
\sim \mu^L \, s_{\cal N}^{\overline{\gamma}_{\cal G} - 1} \,
\overline{{\cal Y}}_{\cal G}\left(\frac{s_1}{s_{\cal N}},
\ldots,\frac{s_{{\cal N}-1}}{s_{\cal N}}
\right),
\end{equation}
with the network exponent 
\begin{equation} \label{nexp_theta}
\overline{\gamma}_{\cal G} = 
1 - d \nu {\cal L} + \sum_{N\ge 1}n_N\overline{\sigma}_N \, \, .
\end{equation}
Overlined symbols refer to polymers at the $\Theta$ point.
In $d = 2$, $\nu = 4/7$ and $\overline{\sigma}_N = (2-N)(2N+1)/42$
\cite{DS87}.

Finally, we note that equation (\ref{nexp}) 
with $\sigma_N = (4 - N^2)/32$ holds for a 
2D dense polymer network which never intersects 
(apart from the vertices).
However, it has been shown recently
that a {\em different}, so--called `Goldstone phase' emerges 
if the strict non-crossing condition is relaxed; 
i.e., if crossings 
{\em are} allowed albeit, disfavoured \cite{JRS03}.
It is argued in reference \cite{JRS03} that 
in this case
the scaling dimensions $X_k$ of the so-called $k$-leg 
operators all vanish. This implies $\sigma_N = 0$ 
in equation (\ref{nexp}) 
(see equation (13) in \cite{Dup86_a} and the 
Euler relations near equation (\ref{euler})
above), i.e., the asymptotic behavior is similar to a network
of chains without self-avoiding constraints 
(ideal polymers). The localisation
exponent $c$ is then simply given by
$c = - \gamma_{\cal G} - m = {\cal L} - m - 1$, 
resulting in
$c$ values of -1, 0, 1, 0, 0, 1, 1 for the 7
contractions shown in figure \ref{fig31}, top 
row from left to right. As argued before, all
contractions with $c < 1$ are delocalised;
in particular the original (uncontracted) trefoil shape,
is the dominant form with the smallest $c$.
Our finding for the 2D dense phase in section \ref{sec_trefoil}
is thus also applicable in the Goldstone phase.
However, in a MC simulation of the Goldstone 
phase it will be hard to identify the knot, 
as additional crossings are allowed and will occur.

\end{appendix}



\begin{thebibliography}{99}
\bibitem{2D} Throughout this paper we employ the 
terms ``2D knot,'' ``2D knot projection,'' and
``2D network'' to denote a polymer chain
confined to two dimensions, which never 
intersects itself apart from explicitly imposed 
``crossings'' (or ``vertices'').
The chain segments adjoining the crossings
are allowed to exchange length with one 
another. Also the slip-linked structures
considered in sections \ref{sec_F8} and 
\ref{sec_trefoil} belong to this class
of systems, as their equilibrium behaviour
is the same. The strict non-crossing 
condition is relaxed only briefly
near the end of appendix \ref{sec_app}, 
where we consider the 2D Goldstone 
phase introduced in reference \cite{JRS03}.
\end{thebibliography}
\end{document}